\documentclass[]{spie}  

\usepackage{amsmath,amsfonts,amssymb}
\usepackage{graphicx}
\usepackage[colorlinks=true, allcolors=blue]{hyperref}

\title{Bringing SOUL on sky}

\author[a]{Enrico Pinna}
\author[a]{Fabio Rossi}
\author[a]{Alfio Puglisi}
\author[a]{Guido Agapito}
\author[a]{Marco Bonaglia}
\author[a]{Cédric Plantet}
\author[a]{Tommaso Mazzoni}
\author[a]{Runa Briguglio}
\author[a]{Luca Carbonaro}
\author[a]{Marco Xompero}
\author[a]{Paolo Grani}
\author[a]{Armando Riccardi}
\author[a]{Simone Esposito}
\author[b]{Phil Hinz}
\author[b]{Amali Vaz}
\author[b]{Steve Ertel}
\author[b]{Oscar M. Montoya}
\author[b]{Oliver Durney}
\author[c]{Julian Christou}
\author[c]{Doug L. Miller}
\author[c]{Greg Taylor}
\author[c]{Alessandro Cavallaro}
\author[c]{Michael Lefebvre}

\affil[a]{INAF - Osservatorio Astrofisico di Arcetri, Italy}
\affil[b]{Steward Observatory, University of Arizona, USA}
\affil[c]{Large Binocular Telescope Observatory, University of Arizona, USA}

\authorinfo{Further author information: \\send correspondence to Enrico Pinna: enrico.pinna@inaf.it}

\begin{document} 
\maketitle

\begin{abstract}
The SOUL project is upgrading the 4 SCAO systems of LBT, pushing the current guide star limits of about 2 magnitudes fainter thanks to Electron Multiplied CCD detector. This improvement will open the NGS SCAO correction to a wider number of scientific cases from high contrast imaging in the visible to extra-galactic source in the NIR. The SOUL systems are today the unique case where pyramid WFS, adaptive secondary and EMCCD are used together. This makes SOUL a pathfinder for most of the ELT SCAO systems like the one of GMT, MICADO and HARMONI of E-ELT, where the same key technologies will be employed. Today we have 3 SOUL systems installed on the telescope in commissioning phase. The 4th system will be installed in a few months. We will present here the results achieved during daytime testing and commissioning nights up to the present date.  
\end{abstract}

\keywords{Pyramid, SCAO, LBT, high contrast, XAO, Adaptive Secondary}

\section{INTRODUCTION}
\label{sec:intro}  
\begin{figure}
    \begin{center}
        \includegraphics[width=0.8\textwidth]{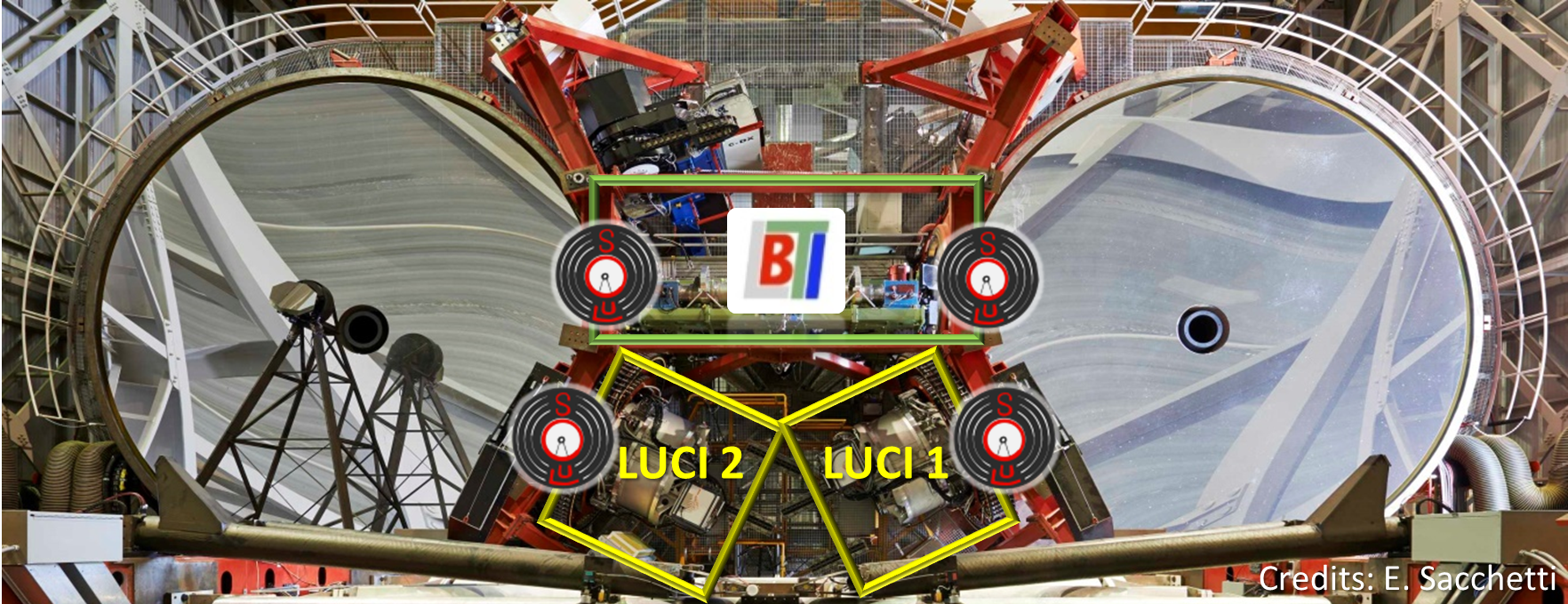}
    \end{center}
	 \caption{\label{fig:LBT} Top view of LBT with yellow lines marking the position of LUCI1 and LUCI2 and green for LBTI. The SOUL logos show the position of the 4 WFSs.}
\end{figure}

The Large Binocular Telescope\cite{2012_Hill_LBT_SPIE} (LBT) is equipped with 4 Single Conjugated Adaptive Optics (SCAO) systems\cite{2018_Christou_SPIE}. All of them are composed by a pyramid WaveFront Sensor (WFS) working with Natural Guide Stars (NGS) and coupled with an Adaptive Secondary Mirror\cite{2010_Riccardi_ASM_SPIE} (ASM) as corrector. Two of these systems\cite{2011_Esposito_FLAO_SPIE} feed two NIR spector-imager (LUCI1 and LUCI2), while the remaining two\cite{2014_Bailey_LBTI-AO_SPIE} feed the focal stations of LBTI. SOUL\cite{2016PINNA_SOUL_SPIE.9909E..3VP} is aimed to upgrade all the 4 systems enabling the AO correction using stars 2 to 3 magnitude fainter. In fig.\ref{fig:LBT} we report a view of LBT with highlighted LBTI and the 2 LUCI together with the position of the 4 SOUL WFSs.

In this work we will report about the current status of the SOUL commissioning (sect.\ref{sect:upgrade}), the calibration of the SOUL-LUCI1 system (sect.\ref{sect:calibration}) and the main result obtained on-sky with LUCI1 and LBTI (sect.\ref{sect:on-sky}).

\section{The upgrade}
\label{sect:upgrade}
\begin{figure}
    \begin{center}
        \includegraphics[width=0.7\textwidth]{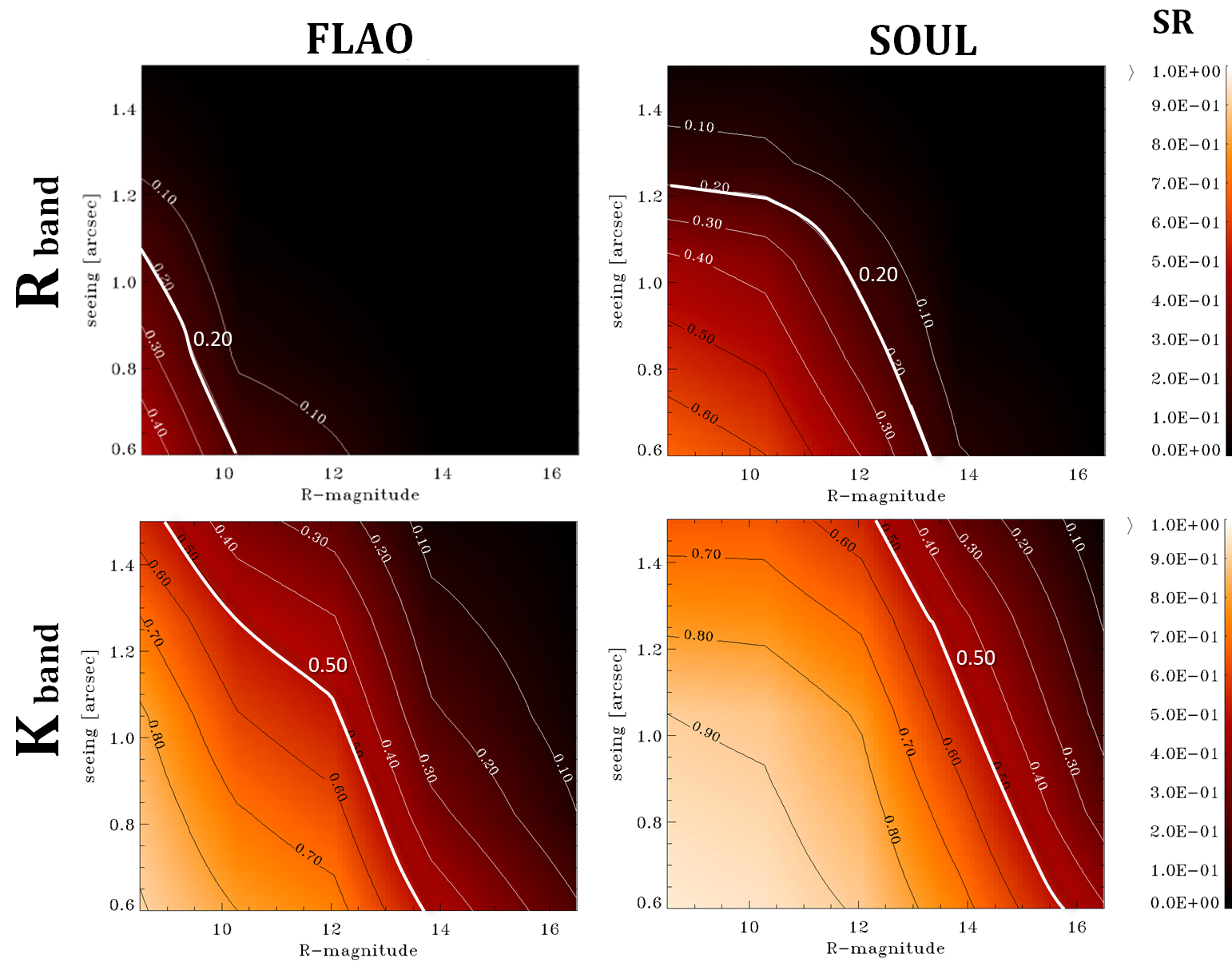}
    \end{center}
	 \caption{\label{fig:E2E} Comparison of the FLAO and SOUL performance in $R$ and $K$-band, as estimated via numerical simulations. The image report the SR value as color scale as function of guide star magnitude (x-axis) and seeing (y-axis). We highlighted in white the line of SR$=20\%$ and $50\%$ for $R$ and $K$-band respectively. }
\end{figure}

The SOUL upgrade has been described already in Pinna+2016\cite{2016PINNA_SOUL_SPIE.9909E..3VP}, where the system details are reported together with the performances estimated via numerical simulations. These have been performed analytically\cite{2019_Agapito_JATIS} and with E2E tools\cite{2016Agapito_PASSATA_SPIE}. Here (fig.\ref{fig:E2E}), we report an update on the expected performance for SOUL in terms of SR values in $R$ and $Ks$-band, compared with those estimated for FLAO under the same conditions. 
Considering the curve of SR=$20\%$ in  $R$-band, we expect that SOUL will deliver diffraction limited images on stars 3 magnitude fainter, moving the current limit from $m_R=10$ to $13$ in good seeing conditions. This is a dramatic increase in the number of target available for SHARK-VIS\cite{2018_MAttioli_SHARK-VIS_SPIE}.
In the $K_s-band$ plot, we can take as reference the line of SR=$50\%$, showing a gain of about 2 magnitudes for good seeings and even larger in poor conditions ($>1.2"$). Being able to deliver SR(K)$>50\%$ with reference star $m_R=15$ and diffraction limited down to $m_R=16$, SOUL is opening to extra-galactic targets, as bright AGN, diffraction limited images and long slit spectroscopy with LUCI2. This is a key feature at LBT, where LGS are available for GLAO correction only\cite{2019Rabien_ARGOS_A&A}.

\begin{figure}
    \begin{center}
        \includegraphics[width=0.6\textwidth]{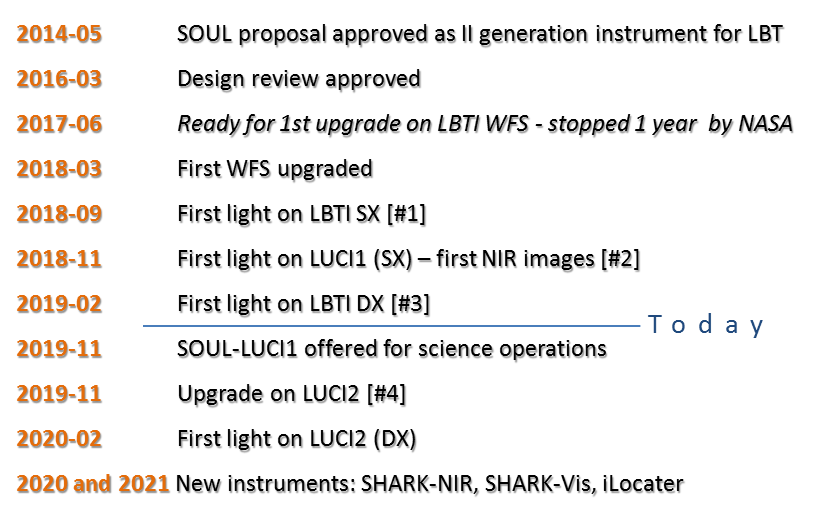}
    \end{center}
	 \caption{\label{fig:timeline}The milestones of the SOUL project.}
\end{figure}

\begin{figure}
    \begin{center}
        \includegraphics[width=0.32\textwidth]{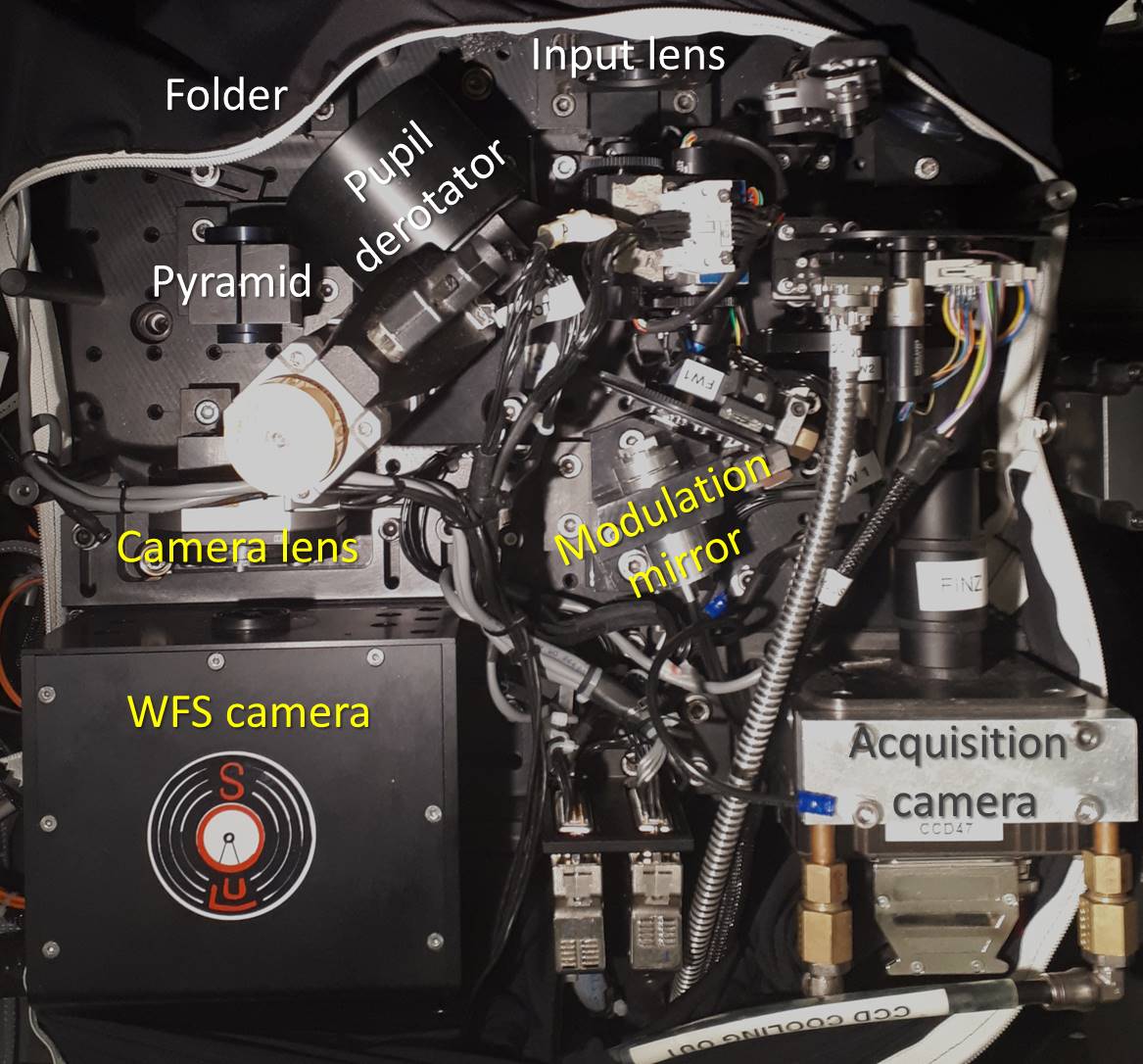}
         \includegraphics[width=0.32\textwidth]{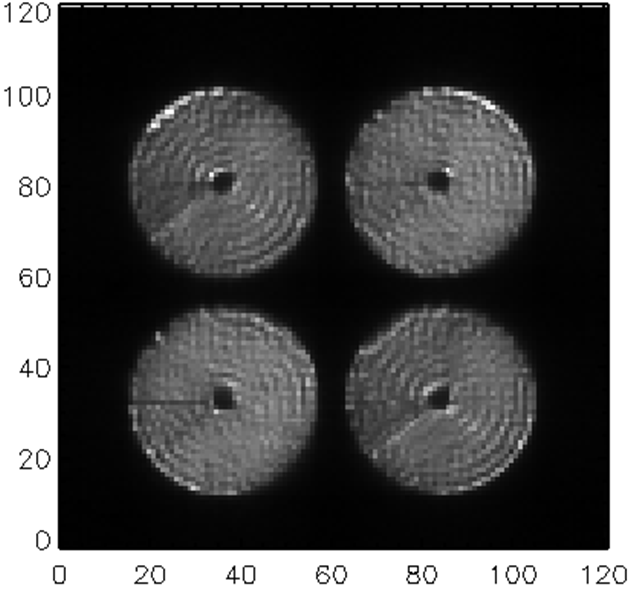}
          \includegraphics[width=0.33\textwidth]{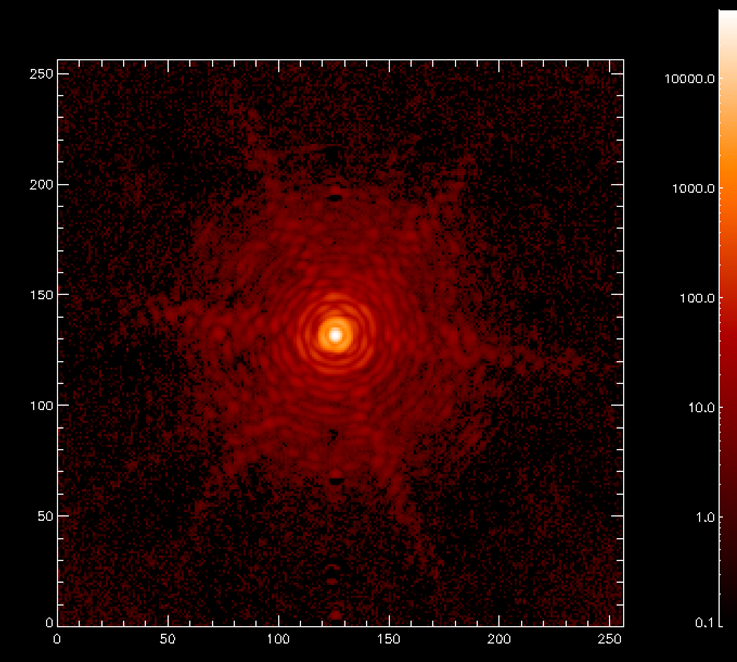}
    \end{center}
	 \caption{\label{fig:WFS-board}\textbf{}{Left}: top view of the SOUL-LUCI1 WFS just before the telescope installation. In yellow the upgraded components. \textbf{Center}: WFS camera frame (averaged over 400 steps) showing the pupil images during on-sky operations. The measurement of the pupil positions confirmed that we achieved the desired magnification ($40\pm1SA$ on the diameter) and separation 
	 ($48.0\pm0.1pix$). At 8 o'clock we can distinguish the beams of the LBT swing arms. \textbf{Right} The first light on SOUL-LUCI1 in November 2018 with a SR(K)=$87\%$ on a reference star $m_R=9.5$ with Seeing $0.7"$, correcting 500 modes at $1.7kHz$ with no NCPA compensation.}
\end{figure}

The project is now in advanced state, as shown in the timeline of fig.\ref{fig:timeline}. After a forced stop in order to allow the completion of the HOST survey\cite{2018_Ertel_HOSTS_AJ}, we upgraded the first WFS (LBTI-SX) on March 2018 and the second one (LUCI1) few months later (fig.\ref{fig:WFS-board}-left). Then, we had the first light for LBTI-SX on September 2018 and for LUCI1 in November of the same year (fig.\ref{fig:WFS-board}-center), delivering the first SOUL image in NIR (fig.\ref{fig:WFS-board}-right). About the WFS upgrade, details are available in the proceedings of this conference, as for tip-tilt mirror calibration\cite{2019ROSSI_SOULTTM_AO4ELT6}, the optical alignment\cite{2019Pinna_pyrAlign_AO4ELT6} and software\cite{2019Rossi_SW4SOUL_AO4ELT6}.   

After the successful first light for the first 2 systems, we upgraded the LBTI-DX system in the winter 2018-2019 having its first light in February. Then the LUCI2 upgraded has been set in stand-by waiting for SOUL-LUCI1 to be operational and offered for routine science observations. The two LUCI \cite{2003_Seifert_LUCIFER_SPIE} \cite{2012_Buschkamp_LUCIintheSky_SPIE} are both facility instruments for LBT and the continuity in science operation is a requirement for the telescope. When we upgraded the FLAO on LUCI1, FLAO-LUCI2 was operative and available for science observations. Then, the first goal has been to provide, on LUCI1, FLAO-like performances with high reliability, releasing it for science observations in SCAO mode and as NGS for ARGOS. This is foreseen to happen in Nov 2019 when LUCI2 FLAO system will then available for the WFS upgrade. The first light on SOUL-LUCI2 in scheduled in February 2020. In the summer 2020 the II generation instruments (SHARK-NIR\cite{2018_Farinato_SHARK-NIR_SPIE}, SHARK-VIS\cite{2018_MAttioli_SHARK-VIS_SPIE} ans iLocater\cite{2016_Crepp_iLocater_SPIE}), all fed by SOUL systems, will start to populate the LBT focal stations.   

\section{System calibration}
\label{sect:calibration}
The calibration of the AO system has been performed at the telescope in daytime using the FLAO source and the retro-reflector at the near focus of the ASM, following the same procedure as for the FLAO systems\cite{2010_Esposito_FLAOCalib_ApOpt}. As calibration we refer to: 1) the Interaction Matrix (IM) measurement;
2) tuning of the AO parameters as function of the reference star brightness. 

The IM measurement has been performed with fast modal push and pull as described in Eposito+2010\cite{2010_Esposito_FLAOCalib_ApOpt}. We measured IMs for the WFS configurations of binning 1x1, 2x2 and 4x4 corresponding to 40, 20 and 10SA on the pupil diameter respectively. The calibration of binning 3x3 has been postponed to the next phase of the project, where we will focus on the ultimate system performances.

The parameter tuning has been done on the SOUL-LUCI1 system, because this instrument can provide $H$-band focal plane images, used here as merit function. On the SOUL-LBTI systems, we ported the same tuning obtained on SOUL-LUCI1. As light source for phase 2), we used the ARGOS calibration unit providing an easier handling with respect to the FLAO one. We applied commands to the ASM mimicking the atmospheric turbulence equivalent to $0.6"$ and $1.0"$ of seeing. We explored binning (1x1, 2x2 and 4x4) and loop frame rates as function of the simulated star brightness. As for the FLAO system, the AO loop gain are optimized automatically at the beginning of each closed loop, scanning the values (on 3 group of modes) and minimizing the WF residuals. So, the gains are not included in the parameter to be tuned. This is the operation baseline, but an on-line optimization of the loop gain mode-by-mode is in progress and we plan to be the final operational mode for SOUL. The on-line optimization of the gains is based on the Genndron-Lena technique applied to the pyramid WFS thanks to the measurement on the WS optical gain as described in Esposito+2015\cite{2015_Esposito_NCPA_AO4ELT4}. 

\begin{figure}
    \begin{center}
        \includegraphics[width=0.6\textwidth]{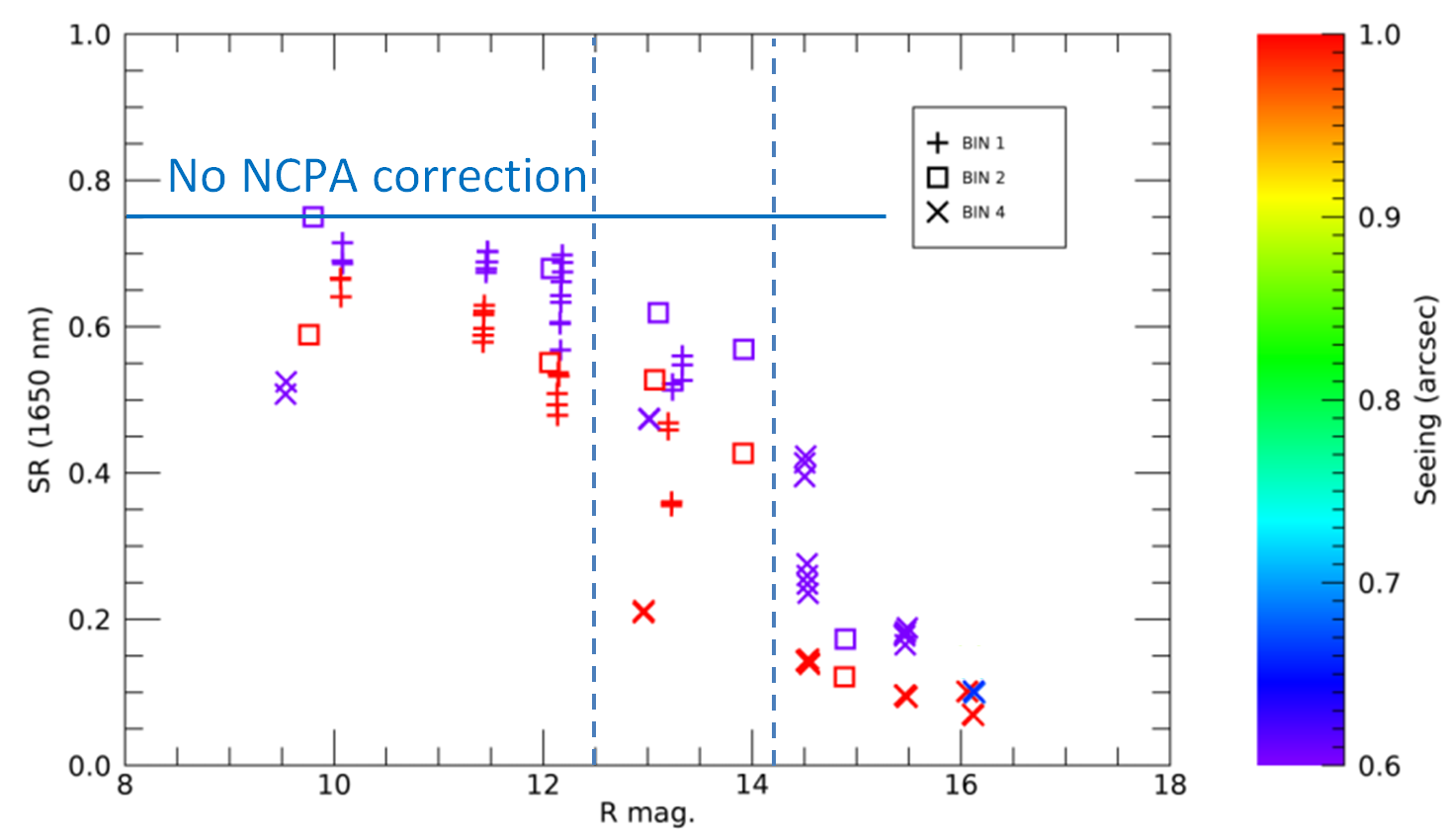}
    \end{center}
	 \caption{\label{fig:tabella_daytime} Long exposure SR measured on LUCI1 during daytime testing with the calibration source. No NCPA correction was applied and the SR values are saturated around $75\%$. Different symbols represent different WFS camera binnings, while colors different seeing values.}
\end{figure}

The numerical simulations indicated that the tip-tilt modulation of $3\lambda/D$ radius is suitable for the full range of magnitude. During this first tuning, we adopted this modulation value as fix, while in the future we plan to optimize it for magnitude and seeing conditions. At this stage, we had no NCPA correction, so the maximum SR in $H$-band was saturated at about $80\%$. The result of these tests is reported in fig.\ref{fig:tabella_daytime}. Table \ref{tab:Tabellone} report the tuning obtained considering the performances measured in fig.\ref{fig:tabella_daytime}. This table is used for the automatic configuration of the system for on-sky operations. The EM gain of the WFS camera is set to $600$ for $m_R>9.5$, the value recommended by First Light, and decreased for brighter magnitudes in order to reduce the excess noise or avoid saturation.

\begin{table}[ht]
\caption{AO parameters tuned on SOUL-LUCI1, as function of the reference star magnitude. These values are used for the automatic configuration of the system for science operations.} 
\label{tab:Tabellone}
\begin{center}       
\begin{tabular}{|l|l|l|l|} 
\hline
\rule[-1ex]{0pt}{3.5ex}  $m_{R}$ & Binning & framerate &  EM gain\\
\hline
\hline
\rule[-1ex]{0pt}{3.5ex}  $<2.5$ & 1x1 &  1.7  & 1\\
\hline
\rule[-1ex]{0pt}{3.5ex}  3.5 & 1x1  & 1.7  & 10 \\
\hline
\rule[-1ex]{0pt}{3.5ex}  4.5 & 1x1  & 1.7  & 20 \\
\hline
\rule[-1ex]{0pt}{3.5ex}  5.5 & 1x1  & 1.7  & 30 \\
\hline 
\rule[-1ex]{0pt}{3.5ex}  6.5 & 1x1  & 1.7  & 100 \\
\hline 
\rule[-1ex]{0pt}{3.5ex}  7.5 & 1x1  & 1.7  & 100 \\
\hline 
\rule[-1ex]{0pt}{3.5ex}  8.5 & 1x1  & 1.7  & 100 \\
\hline 
\rule[-1ex]{0pt}{3.5ex}  9.5 & 1x1  & 1.7  & 300 \\
\hline 
\rule[-1ex]{0pt}{3.5ex} 10.5 & 1x1  & 1.7  & 600 \\
\hline 
\rule[-1ex]{0pt}{3.5ex} 11.5 & 1x1  & 1.25  & 600 \\
\hline 
\rule[-1ex]{0pt}{3.5ex} 12.5 & 1x1  & 0.75  & 600 \\
\hline 
\rule[-1ex]{0pt}{3.5ex} 13.0 & 2x2  & 1.0  & 600 \\
\hline 
\rule[-1ex]{0pt}{3.5ex} 14.0 & 2x2  & 0.8  & 600 \\
\hline
\rule[-1ex]{0pt}{3.5ex} 14.5 & 4x4  & 1.2  & 600 \\
\hline 
\rule[-1ex]{0pt}{3.5ex} 15.5 & 4x4  & 0.5  & 600 \\
\hline 
\rule[-1ex]{0pt}{3.5ex} 16.5 & 4x4  & 0.2  & 600 \\
\hline 
\rule[-1ex]{0pt}{3.5ex} $>17.5$ & 4x4 & 0.1  & 600 \\
\hline 
\end{tabular}
\end{center}
\end{table}

We have to mention that the AO control loop at binning 1x1 adopts the "forgetting factors", replacing the classical pure integrator, as detailed in these proceedings\cite{2019_Agapito_Goldfish_AO4ELT6}. This control demonstrated to relax significantly the forces on the ASM, when high orders are controlled, producing a remarkable improvement of the loop robustness. In this phase of the project, we were limited to a maximum of 500 controlled modes due to the current optical calibration of the ASM\cite{2018_Briguglio_ASmCalibra_NatSR}. The SX-ASM calibration is 4 years old and does not match anymore the response of the optical surface at the higher spatial frequencies. In order to achieve the ultimate performance of SOUL, a new calibration of the ASM with the interferometer is planned in the next months. As you can notice in table \ref{fig:tabella_daytime}, the maximum frame rate is set to $1.7kHz$, while the goal for SOUL is $2.0kHz$. This is due to a limitation in the fast diagnostic and could be solved in the next future. However, at this stage, the frame rate of $1.7kHz$ is not the limiting factor for the SOUL performances. 

As future work in the daytime calibrations, we have: calibration of WFS binning 3x3 (improving in the regime around $m_r=14.5$), trim the number of modes as function of the magnitude and optimize frequencies at the faint end ($m_R>16$).

%
%


\section{On-sky results}
\label{sect:on-sky}
In this section, we report the results obtained on the system up to June 2019. On-sky commissioning is ongoing and just a fraction of the planned nights (5 of 15) has been executed, mainly due to the weather conditions really poor in winter 2018-2019. 

\subsection{LUCI1}
\begin{figure}
    \begin{center}
        \includegraphics[width=0.6\textwidth]{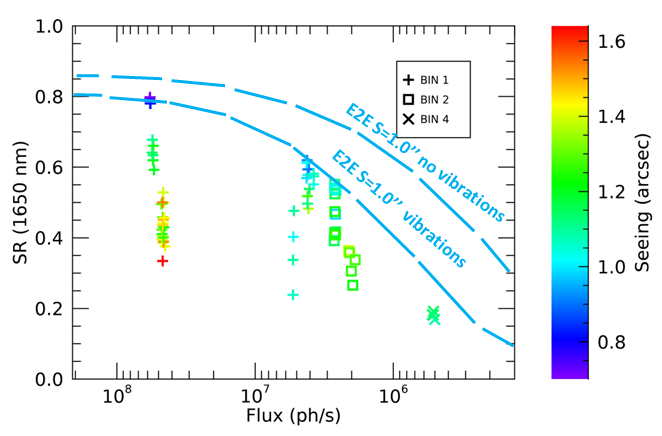}
    \end{center}
	 \caption{\label{fig:SR_onsky} SR values at $1650nm$ reported as function of the flux detected on the WFS and the Dimm seeing in the line of sight (colors). Different symbols represent different WFS samplings (binnings). The dashed lines report the simulations values expected with no and strong telescope vibrations for a seeing of $1.0"$.}
\end{figure}

In fig.\ref{fig:SR_onsky} we report SR values (at 1650nm, measured at this wavelength or rescaled using Marechal's approximation) obtained on LUCI1 during the commissioning at different star fluxes and seeing conditions. All these values are long exposures (10 to 60s), obtained as direct sum of sub-frames with no shift and add. We mainly focused on the higher part of the flux range (binning 1x1 and binning 2x2), while the faint end is still to be explored. We experienced seeing usually higher than $0.9"$ on the line of sight. In the plot we report as comparison the values expected from simulation with seeing of $1.0"$ for two cases, considering no and strong telescope vibrations. The correction of NCPA was not calibrated at that time and so no correction was applied. With LUCI N30 camera NCPA limits the maximum achievable SR around $80\%$ in $H$-band. 

\begin{figure}
    \begin{center}
        \includegraphics[width=0.9\textwidth]{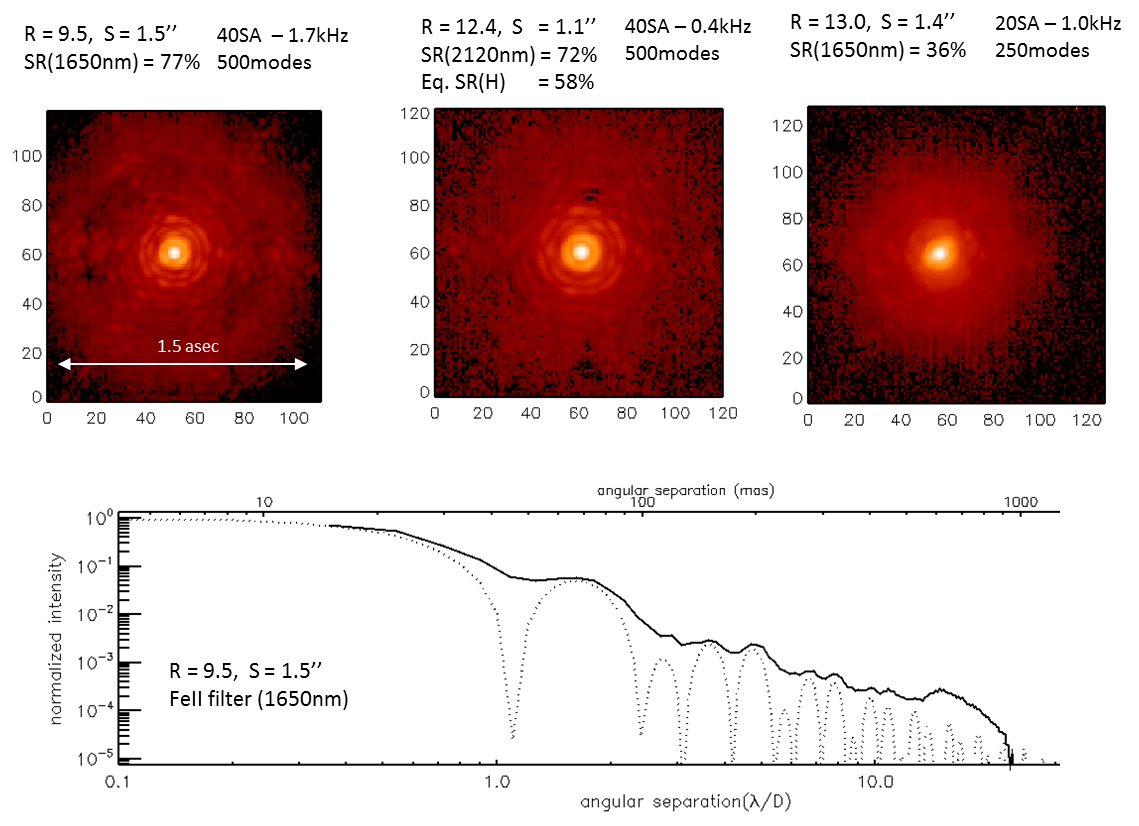}
    \end{center}
e	 \caption{\label{fig:BestPSF} \textbf{Top:} a sample of PSF measured on LUCI1 during the SOUL commissioning. All are long exposures on the AO reference star. \textbf{Bottom:} radial profile of the top-left PSF (solid line) compared with the theoretical PSF on the LUCI1 focal plane (dotted). The plot shows a raw contrast of the order of $10^{-4}$ at $400-500mas$ off the peak.}
\end{figure}

In this first phase of commissioning, we collected good performances showing already the gain of SOUL w.r.t. to FLAO. In fig.\ref{fig:BestPSF}, we report a sample of PSFs measured during the commissioning nights. In fig.\ref{fig:BestPSF} top-left we show the long exposure PSF in bright regime ($m_r=9.5$) under strong seeing conditions (fast variability $1.2"-1.5"$), showing a SR of $77\%$.
This results demonstrates the robustness of SOUL in high seeing condition, allowing to correct 500 modes, while FLAO was forced to reduce the number of modes to 250 in order to limit the forces on the ASM. This gain is due to the higher frame rate and to the control with forgetting factors\cite{2019_Agapito_Goldfish_AO4ELT6}. In fig.\ref{fig:BestPSF} top-center we show the PSF of a guide star of $m_R=12.4$ with seeing $1.1"$. Here SOUL can still use 40x40SA (500 modes), while FLAO, due to the detector noise, requires the sampling reduction to 15x15SA, with a maximum of 150 controlled modes. The measured long exposure SR(K) of $72\%$  is in line with the expected one (SOUL simulations $74\%$), confirming the expected gain on FLAO (simulations $47\%$). The last example of PSF we report is in fig.\ref{fig:BestPSF} top-right, where we obtained $SR(H)=36\%$ on a reference star of $m_R=13.0$ with pupil sampling 20x20 and 250 modes correcting a seeing of $1.4"$. Again, the measured SR is in good agreement withe the expectation of SOUL's simulations ($34\%$), clearly overtaking FLAO (simulation in same conditions: SR(H)=$12\%$).


\subsection{LBTI} 
 Both systems are operational and routinely used for science since their respective upgrades. The commissioning of both systems has been carried on to the point where system can be operated by the LBTI team, enabling science operation both in imaging and interferometric modes. The automation of the AO operations is progressing in the commissioning of SOUL-LUCI1 and all improvements are periodically ported to LBTI systems. This is now allowed by the alignment of the control SW on all the 4 SOUL systems. The performance tuning and their quantitative assessment is postponed to the availability of SHARK-NIR and SHARK-Vis that will provide PSF images at shorter wavelength in daytime with the calibration source.

Said that, the qualitative feedback of the LBTI observers states that SOUL shows better stability (higher efficiency in time) and increased limiting magnitude. In addition, the system can be operated and delivers good correction under higher seeing ($>1.2"$), which is critical for successful, routine AO operations on Mt. Graham given the often relatively poor seeing compared to sites of other 8m telescopes.

\begin{figure}
    \begin{center}
        \includegraphics[width=0.22\textwidth]{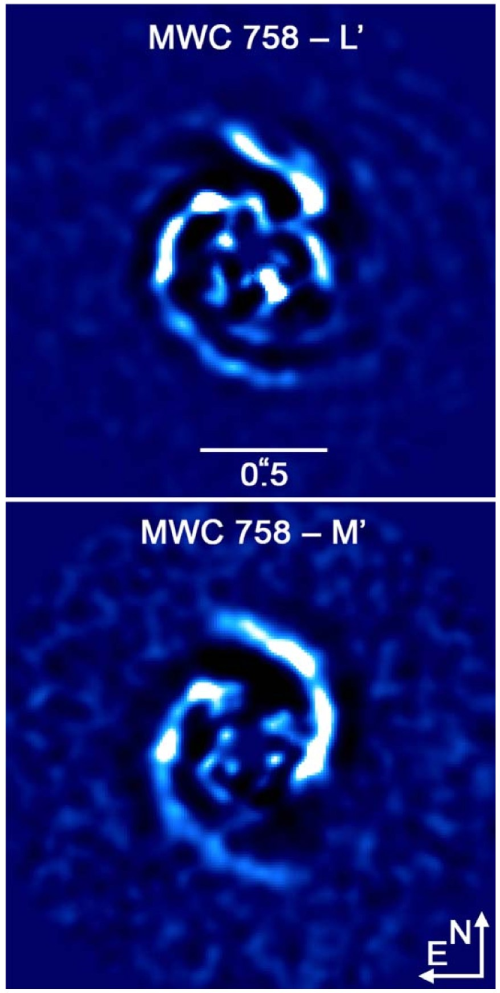}
    \end{center}
	 \caption{\label{fig:arms}  Example of observation with SOUL-LBTI on a bright source. Here a young star shows spiral arms in $L'$ and $M'$-band in a dual-aperture observation just after the SOUL first light on LBTI-DX.}
\end{figure}

\begin{figure}
    \begin{center}
        \includegraphics[width=0.8\textwidth]{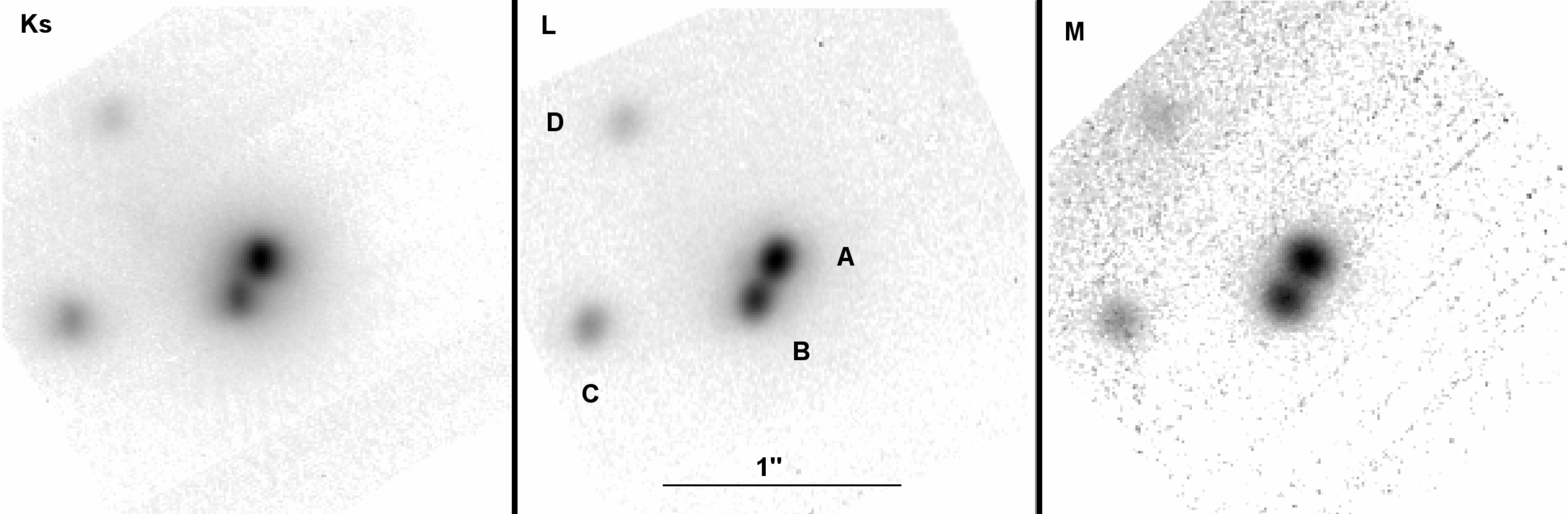}
    \end{center}
	 \caption{\label{fig:QSO}  Quad lensed quasar observed with SOUL-LBTI SX on this February. From left to right the images are in $K_s$, $L$ an $M$-bands. This images allowed to compare the flux ration of the central peaks with the model. Credit:  T. J. Jones \& L. L. R. Williams, University of Minnesota.}
\end{figure}

As example of the scientific production of the LBTI SOUL systems, we report here two cases, one with bright and one with fainter AO guide star. Closing the AO loop on the target as reference ($m_r=8.3$), the images of young star MWC 758 shows the first detection of its spiral arms in $M'$ and the deepest one in $L'$ (see fig.\ref{fig:arms}). All the details and scientific results have been published in Wagner+2019\cite{2019_Wagner_QSO}. On the fainter end, we report the observation of a quad lensed quasar of integrated magnitude over the 4 components of $m_R=15.3$. Again guiding on the target, well resolved images have been acquired in $K_s$, $L'$ and $M'$-bands (see fig.\ref{fig:QSO} and Jones+2019\cite{2019_Jones_quadLens_AJ}).

The upgrade AO has also been demonstrated to be usable for nulling interferometry, and a first data set to characterize the impact of the upgrade on the nulling data quality has been obtained (to be analyzed). Higher stability and wavefront correction have a positive impact on the efficiency of nulling observations, the range of suitable seeing conditions, and the ability to observe at low elevations. The actual nulling data quality is thought to be limited by other factors, so that only minor improvements from the SOUL upgrade are expected here.

\section{Conclusions}
The SOUL project is in progress and in advanced status, having upgraded 3 of 4 systems. The two serving LBTI are routinely used for science operations both in imaging and interferometric modes. The one serving LUCI1 will be released for science on November 2019. The fourth system will be upgraded in the winter 2019-2020. 

The on-sky performances with LUCI1 still requires a full characterization under the full range of reference star magnitudes and seeing conditions. However, the first results show that the system is able to reach the expected performances in terms of SR in the range $9<m_R<13.5$ and robustness to strong seeing ($>1.2"$). The next semester of commissioning will focus on the performance assessment and optimization. The final phase of this activity will happen when SHARK-VIS and SHARK-NIR will be coupled to the SOUL system delivering high contrast images (2020-2021).

SOUL is currently the only system assembling together a pyramid WFS equipped with an EMCCD detector and an adaptive secondary mirror as corrector. This combination allows to push at the ultimate limit the SCAO NGS systems, fully exploiting the photon flux from faint stars. The same approach will be adopted in many of the SCAO systems of the ELTs, as those for MICADO\cite{2018_Clenet_MICADO-SCAO_SPIE} and HARMONI\cite{2018_Dohlen_HARMOMY-AO_SPIE} on the E-ELT and the NGAO\cite{2018_Bouchez_GMTAO_SPIE} of GMT. Therefore, SOUL represents a unique pathfinder for such systems that will provide the first AO correction on ELTs.

\acknowledgments 
 
The SOUL team wants to acknowledge the valuable and constant support provided by the LBTO personnel during the integration and commissioning activity both at the Steward Observatory labs. and on the mountain.

\bibliography{report} 
\bibliographystyle{spiebib} 

\end{document}